\documentclass[a4paper,11pt]{iopart}

\pdfoutput=1 

\usepackage{upgreek}                    
\usepackage{color}
\usepackage{url}

\usepackage{amssymb,amsfonts,graphicx}

\newcommand{\micron}{\, \rm{\upmu m}}
\newcommand{\erf}{\textrm{erf}}
\newcommand{\rchi}{{\mathpalette\irchi\relax}}
\newcommand{\irchi}[2]{\raisebox{\depth}{$#1\chi$}}

\begin{document}
\title[]{Nuclear emulsions for the detection of micrometric-scale fringe patterns: an application to positron interferometry}

\author{S. Aghion$^{a,b}$, A. Ariga$^{c}$, M. Bollani$^{d,b}$, A. Ereditato$^{c}$, \\R. Ferragut$^{a,b}$, M. Giammarchi$^{b}$, M. Lodari$^d$, C. Pistillo$^c$, \\ \footnote{Corresponding author}S. Sala$^{e,b}$, P. Scampoli$^{c,f}$, M. Vladymyrov$^c$}

\address{$^a$LNESS Laboratory and Dipartimento di Fisica, Politecnico di Milano, Via Anzani 42, 22100 Como, Italy}
\address{$^b$Istituto Nazionale di Fisica Nucleare, Sezione di Milano, via Celoria 16, 20133 Milano, Italy}
\address{$^c$Albert Einstein Center for Fundamental Physics, Laboratory for High Energy Physics, University of Bern, Sidlerstrasse 5, 3012 Bern, Switzerland}
\address{$^d$Istituto di Fotonica e Nanotecnologie del CNR, LNESS Laboratory, via Anzani 42, 22100 Como, Italy}
\address{$^e$Dipartimento di Fisica "Aldo Pontremoli",  Universit\`{a} degli Studi di Milano, via Celoria 16, 20133 Milano, Italy}
\address{$^f$Dipartimento di Fisica "Ettore Pancini", Universit\`{a} di Napoli Federico II, Complesso Universitario di Monte S. Angelo, 80126 Napoli, Italy}

\ead{simone.sala@mi.infn.it}
\begin{abstract}
Nuclear emulsions are capable of very high position resolution in the detection of ionizing particles. This feature can be exploited to directly resolve the micrometric-scale fringe pattern produced by a matter-wave interferometer for low energy positrons (in the 10-20 keV range). We have tested the performance of emulsion films in this specific scenario. Exploiting silicon nitride diffraction gratings as absorption masks, we produced periodic patterns with features comparable to the expected interferometer signal. Test samples with periodicities of 6, 7 and 20 $\rm{\upmu m}$ were exposed to the positron beam, and the patterns clearly reconstructed. Our results support the feasibility of matter-wave interferometry experiments with positrons.
\end{abstract}

\clearpage

\section{Introduction}
\label{sec:introduction}
In the framework of the QUPLAS (QUantum interferometry and gravitation with Positrons and LASers) experiment \cite{amattint1,amattint2,emulsions_old}, we suggested that emulsion detectors are a promising option to carry out matter-wave interferometry experiments with low energy positrons. We foresee to employ micrometric diffraction gratings with unequal periodicities in a generalized period-magnifying Talbot-Lau \cite{lau,talbotlau} configuration (see \cite{amattint2} for more details), designed to operate on a $10-20 \, \rm{keV}$ positron beam. In particular, gratings with a nominal periodicity of $1 \micron$ and $1.2 \micron$ can be combined to produce interference fringes with a $d=6 \micron$ periodicity \cite{amattint2}. The total length of this layout is approximately $70 \, \rm{cm}$, at resonance for $14 \, \rm{keV}$ positrons; the positron energy $E$, hence their de Broglie wavelength $\lambda = h / \sqrt{2 m E}$ \cite{debroglie} can then be changed, and the contrast of the fringes measured as a function of $\lambda$. A modulation in contrast with $\lambda$ is a clear signature of genuine wave-like quantum interference, as opposed to geometrical shadow (moir\'{e}) effects \cite{amattint1}. A hit position resolution for positrons significantly better than $1 \micron$ is required on the detector plane to resolve a $6 \micron$ pattern with good contrast. This resolution is achieved by emulsion detectors which feature an intrinsic position resolution at the level of $0.1 \micron$ \cite{aegis2} and were already employed for similar studies in the past \cite{nature}.
The sinergy of a high resolution detection and a detailed theoretical model of the signal \cite{croninmodel} could allow to extract more information from  a matter-wave interferometer with respect to the commonly used approach of moving grating masks in a three-grating interferometer \cite{three-grating}.

One of the main challenges in positron interferometry (or antimatter in general), is to produce an intense beam with suitable spatial coherence and monochromaticity to obtain high contrast fringes. Therefore, the detector for the interferometric signal must be able to operate reliably for long data taking runs (of the order of several days) and to consistently reconstruct fringe patterns on a relatively large surface (tens of $\mathrm{mm}^2$). This should be possible even when the combination of statistics and contrast is such that the pattern is not recognizable by visual inspection or simple analysis techniques. 
In this paper we provide experimental evidence that the emulsion-based detector specifically developed for this application satisfies both requirements. 
This claim is supported by measurements of periodic fringe patterns obtained by means of silicon nitride diffraction gratings able to produce a signal comparable to the one expected from the QUPLAS interferometer. The large-area free standing membranes were indeed placed in close contact with the emulsion surface to act as simple absorption masks, and exposed to the positron beam of the L-NESS laboratory in Como (Italy). The emulsion were then analysed by the automated microscopes at the emulsion laboratory of the University of Bern. For the analysis of emulsion data we developed two pattern recognition techniques, which can be readily adapted to the to the search for periodic fringes produced by matter-wave interferometers.

\section{Materials and methods}
\label{sec:matmethods}

Nuclear emulsions are composed of silver-bromide microcrystals with a diameter of approximately $0.2 \micron$, embedded in a gelatin matrix. The latent image left by the passage of ionizing particles is developed via a chemical process to silver grains approximately $1 \micron$ in diameter, visible through an optical microscope (see \cite{emulsion_technology} for a review of the technology).
The emulsion gel used in this work was produced at the Nagoya University in Japan. It features a content of silver bromide crystals of approximately $55 \%$ in volume, and a low background of thermally induced grains, at the level of $1-2 \, \mathrm{grains} / 1000 \micron^3$ \cite{emulsions_old}. This gel is then poured on a glass plate at the Laboratory of High Energy Physics of the University of Bern. The glass support features a thermal expansion coefficient about one order of magnitude smaller than the usual plastic supports for emulsions, to ensure the necessary stability to detect micrometric scale fringe patterns covering a surface of some $\mathrm{mm}^2$. A $(1.0 \pm 0.1) \micron$  thick gelatin layer is applied on the active emulsion surface, to protect it from mechanical stress during handling and transportation, which would result in an increased background. Glycerine is also added to the gel at a concentration of $1.5 \%$ to allow operation in vacuum \cite{emulsions_in_vacuum}. 

\begin{figure}[htbp]
\centering 
\frame{\includegraphics[width=.45\textwidth]{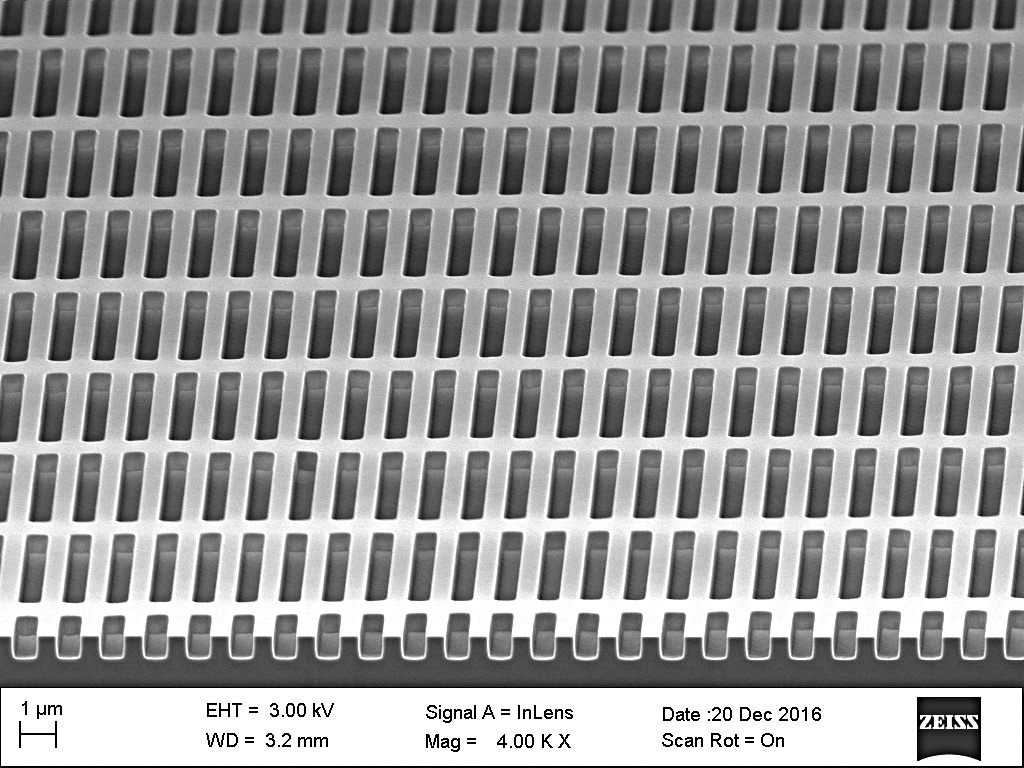}}
\qquad
\frame{\includegraphics[width=.45\textwidth,trim={0 0 0 0.6cm},clip]{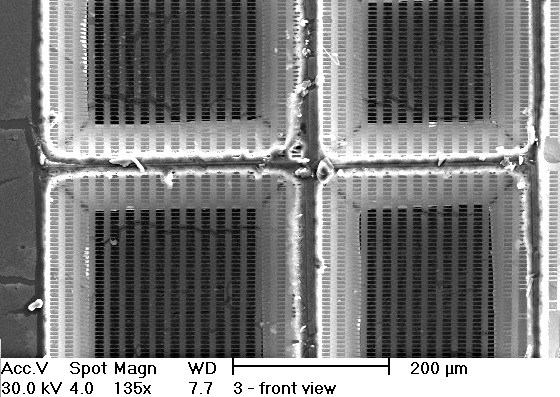}}
\caption{\label{fig:gratings} SEM images of Grating A (left) and Grating B (right).}
\end{figure}

Two different silicon nitride diffraction gratings were exposed to the positron beam:
\begin{itemize}
\item \textbf{Grating A} was produced by LumArray Inc., specifically for the Talbot-Lau positron interferometer. The transmissive region is a free standing silicon nitride membrane, approximately $ 600 \, \mathrm{nm}$ thick, coated on both sides with a $10 \, \mathrm{nm}$  thick gold layer to prevent charge build-up. The membrane has a surface of $3 \times 3 \, \mathrm{mm}^2$, and is patterned with a system of rectangular apertures. Slits with a periodicity $d_1 = (1209.7  \pm 0.3) \, \rm{nm}$ and an approximately $50 \%$ open fraction run along the horizontal direction (see figure \ref{fig:gratings}). In the orthogonal direction, the structure has a period $d_2 = ( 7000 \pm 50 ) \, \rm{nm}$ and an open fraction of $(79 \pm 3) \%$. 
\item \textbf{Grating B} was produced in the framework of our collaboration at the L-NESS laboratory in Como (Italy). This free standing silicon nitride membrane ($325 \, \mathrm{nm}$ thick, coated with a $10 \, \rm{nm}$ gold layer) spans several adjacent but disconnected regions, each one $300 \times 300 \micron^2$ wide. A system of slits is patterned in two orthogonal directions, in analogy with the previous case. Periods are $d_3= (6000.0 \pm 0.5) \, \rm{nm}$ (open fraction of $(58 \pm 3) \%$) and $d_4 = (20000.0 \pm 0.5) \, \rm{nm}$ (open fraction of $(54 \pm 1)\%$). 
\end{itemize}
\begin{figure}[htbp]
\centering 
\includegraphics[width=.5\textwidth]{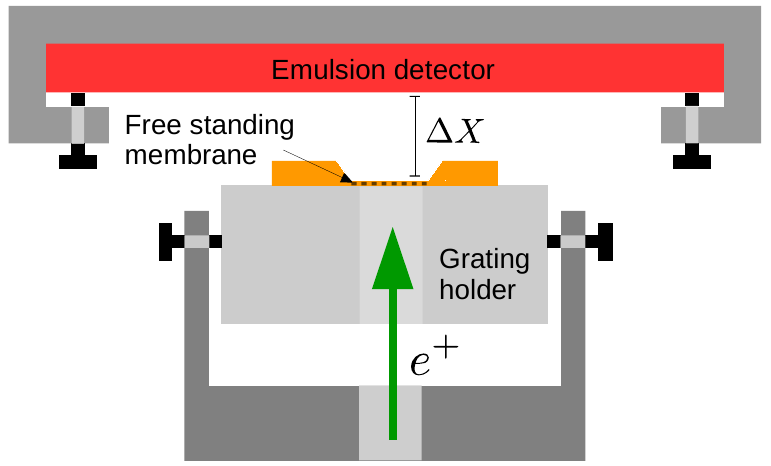}
\caption{\label{fig:mechanical_system} Sketch of the system used to put the grating membranes in close contact with the emulsion detector. The grating holder was mounted on an $x-y$ translation stage (not shown), to allow alignment of the grating membrane with the beam. The membrane-to-emulsion distance is defined as $\Delta X$.}
\end{figure}

The gratings were positioned in close contact with the emulsion surface. This was achieved by means of a carefully aligned grating holder (see figure \ref{fig:mechanical_system}), with the necessary degrees of freedom for the grating to seat flat on the emulsion surface. In the case of Grating A, an arbitrarily small distance was possible. However a spacing of $\Delta X_{A} \approx 0.1 \, \mathrm{mm} $ was set to avoid breakage of the thin membrane. For sample B, the membrane-to-emulsion distance was limited by the thickness of the silicon wafer, which was mounted as shown in figure \ref{fig:mechanical_system}. Therefore we estimated a distance of the order of $\Delta X_{B} \approx 0.3  \, \mathrm{mm}$.

The grating/emulsion system was exposed to the monoenergetic continuous positron beam of the L-NESS laboratory. Positrons are emitted by a $\small ^{22}$Na source and moderated by a monocrystalline tungsten foil. Their energy is then controlled by means of an electrostatic system, from $0.1 \, \mathrm{keV}$ to $20 \, \mathrm{keV}$. The beam operated at a pressure in the $10^{-6} - 10^{-7} \mathrm{mbar}$ range, and delivered a nearly Gaussian beam spot of approximately $ 2.3 \, \mathrm{mm}$ FWHM on the grating plane with a flux of approximately $5 \times 10^3 \, s^{-1}$. The grating membrane was centered with respect to the peak beam intensity by means of a reference laser.

\section{Pattern detection techniques}
\label{sec:techniques}

\begin{figure}[htbp]
\centering 
\includegraphics[width=.95\textwidth]{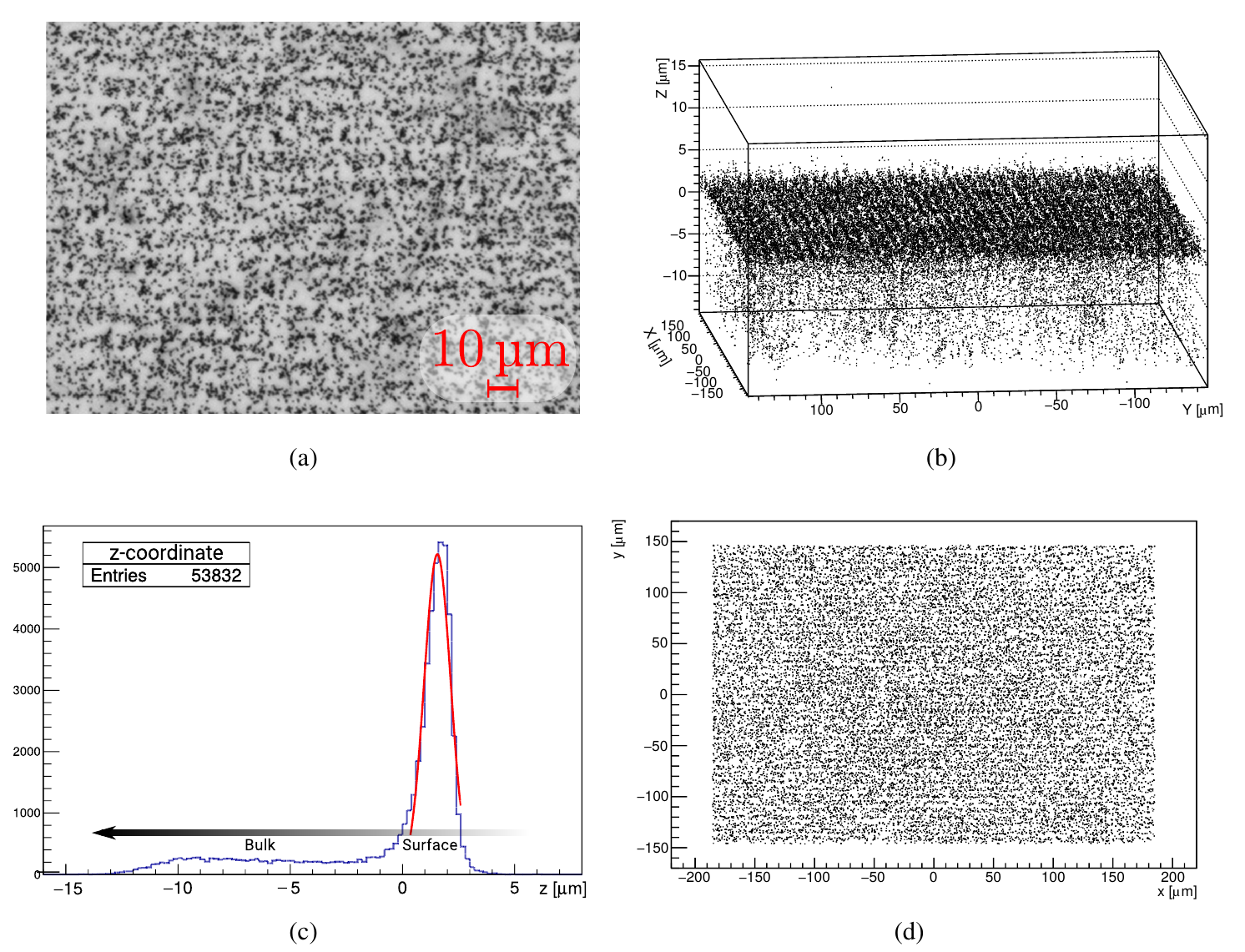}
\caption{\label{fig:analysis_fromview} Center portion of a raw microscope frame at fixed $z$ position (a), $(x,y,z)$ position of the reconstructed clusters in the view (b), histogram of the $z$ coordinate of the clusters and Gaussian fit used for the selection procedure (c). The arrow indicates the direction of incoming positrons. Note than the origin of the $z$-axis is arbitrary and different for each view; the cut is determined independently for each single view. Plot of the $(x,y)$ position of the clusters (d). All the above images refer to the same view, from the exposure of Grating A. }
\end{figure}

At the emulsion scanning facility of the University of Bern, optical microscopes can perform an automatic scan of emulsion surfaces. The microscope camera is equipped with a $1280 \times 1024$ pixels CMOS sensor and grabs images corresponding to $378 \times 294 \micron^2$ emulsion surface. The center portion of one of these images selected from exposure of grating A is shown in figure \ref{fig:analysis_fromview} (a).
A positioning stage moves the microscope on the horizontal ($x-y$) plane; for each given position a sequence of 35 images is taken by shifting the focal plane vertically along the $z$-axis (the pitch between consecutive images is $1.5 \micron$). The silver grains are reconstructed as clusters by a specific algorithm which assigns them their corresponding $(x,y,z)$ coordinates in the reference frame of the stage. In figure \ref{fig:analysis_fromview} (b) the distribution of the cluster coordinates from all the images taken at a given horizontal position is shown, and in the following we will refer to these sets of data as \emph{views}. The volume of interest is scanned by repeating this procedure for adjacent views.

Positrons in the $E<20 \, \mathrm{keV}$ energy range can penetrate less than $1 \micron$ in the active  volume of the emulsion detector before annihilation takes place, therefore a sharp peak close to the surface is a clear signature of positron annihilation grains, as it is clearly visible from the histogram of figure \ref{fig:analysis_fromview} (c).

Background grains detected in the emulsion are due to various sources: an intrinsic noise due to thermal effects, cosmic rays and Compton electrons from the $511 \, \rm{keV}$ gammas produced in positron annihilation events.

To isolate the signal, a Gaussian fit is performed on the peak of the histogram and a cut on the z coordinate is made using the standard deviation $\sigma_z$ and mean $z_0$ from the fit results, namely $\left| z_0 -z \right| < 1.5 \sigma_z $.

Close to the edge of the microscope frames, optical aberrations are not negligible. The decrease in sharpness and contrast in that region could worsen the performance of the clustering algorithm, therefore a cut is also performed in $(x,y)$, to an area of $340 \times 270 \micron^2$. Furthermore, aberrations result in a slight curvature of the average positron implantation depth over the surface of the view, artificially enlarging the width of the peak in $z$. The $(x,y,z)$ coordinates are thus fit with a polynomial of the form $f(x,y) = (x-a)^2 + (y-b)^2  + c$, and the $z$ positions corrected accordingly. The goal of this correction is to improve the effectiveness of the cut on $z$, and has indeed been found to consistently improve the measured contrast. Therefore the $(x,y)$ position is not altered (this would require a more detailed model of the distortion profile, and is out of the scope of this work). The resulting $(x,y)$ distribution after aberration correction and the application of the above mentioned selection criterion on $z$ coordinate is shown in figure \ref{fig:analysis_fromview} (d). %where a hint of a periodic structure (with the expected $7 \micron$ periodicity) is appreciable even by visual inspection.

The analysis of emulsion data needs to take into account the rotation angle $\alpha$ between the laboratory and the microscope reference frames. Therefore, the periodic signal is expected in a linear combination of the $(x,y)$ coordinates such as $t=-x \sin \alpha + y \cos \alpha$, with the rotation angle $\alpha$ to be determined.

The intensity profile generated by the periodic absorption mask is well approximated by a \emph{square wave}. Specifically for a period $d$, and an open fraction $f_o$, neglecting an overall normalization constant, it can be written as follows:
\begin{equation}
I(t) = \sum_{k=-\infty} ^{\infty} \rchi_{\left[ kd , (f_o + k)d \right]}(t) + A(C_0) \rchi_{\left[ (f_o + k)d , (1+k)d \right]}(t),
\label{eqn:ideal_signal}
\end{equation}

\noindent where $\rchi_{\left[a,b \right]} (t)$ is the characteristic function of the interval $[a,b]$, and the parameter $A(C_0) <1$ is related to the contrast by $A(C_0) = (1 -C_0)/(1+C_0)$. The contrast $C_0$ depends on the energy dependent positron transmission rate through the closed portion of the grating periodic structure and the protective emulsion layer, and represents the contrast of the signal before the introduction of finite resolution effects. It can be estimated by a Monte Carlo simulation with the \emph{PENELOPE} (Penetration and ENErgy LOss of Positrons and Electrons) package \cite{penelope,pypen}. 

The effective resolution of the emulsion-grating system, defined as a Gaussian smearing of the $(x,y)$ position of the clusters is given by three main contributions: the intrinsic resolution of the scanning and clustering procedure $\sigma_{\rm{INT}} \approx 0.1 \micron$, the multiple scattering in the emulsion protective layer ($\sigma_{\rm{MSL}} \approx 0.3 \micron$) \footnote{Estimated with a Monte Carlo simulation in the relevant energy range for these exposures, given the density and composition of the protective layer \cite{emulsions_old}. We do not give a detailed discussion of the error, which is dominated by the $10\%$ uncertainty on the thickness of the layer.}, and finally $\sigma_{\rm{DIV}} = \sigma_{\theta} \Delta X$. The latter is due to the angular divergence of the beam, projected through the emulsion-membrane distance $\Delta X$. The beam divergence was estimated to be $\sigma_{\theta} \approx 6 \, \rm{mrad}$, by measuring its diameter on two planes $65 \, \mathrm{cm}$ apart, one of which is the beam waist. This makes $\sigma_{\rm{DIV}}$ the dominant contribution to the total resolution already for $ \Delta X > 0.05 \, \rm{mm}$.

\begin{figure}[htbp]
\centering 
\includegraphics[width=.75\textwidth]{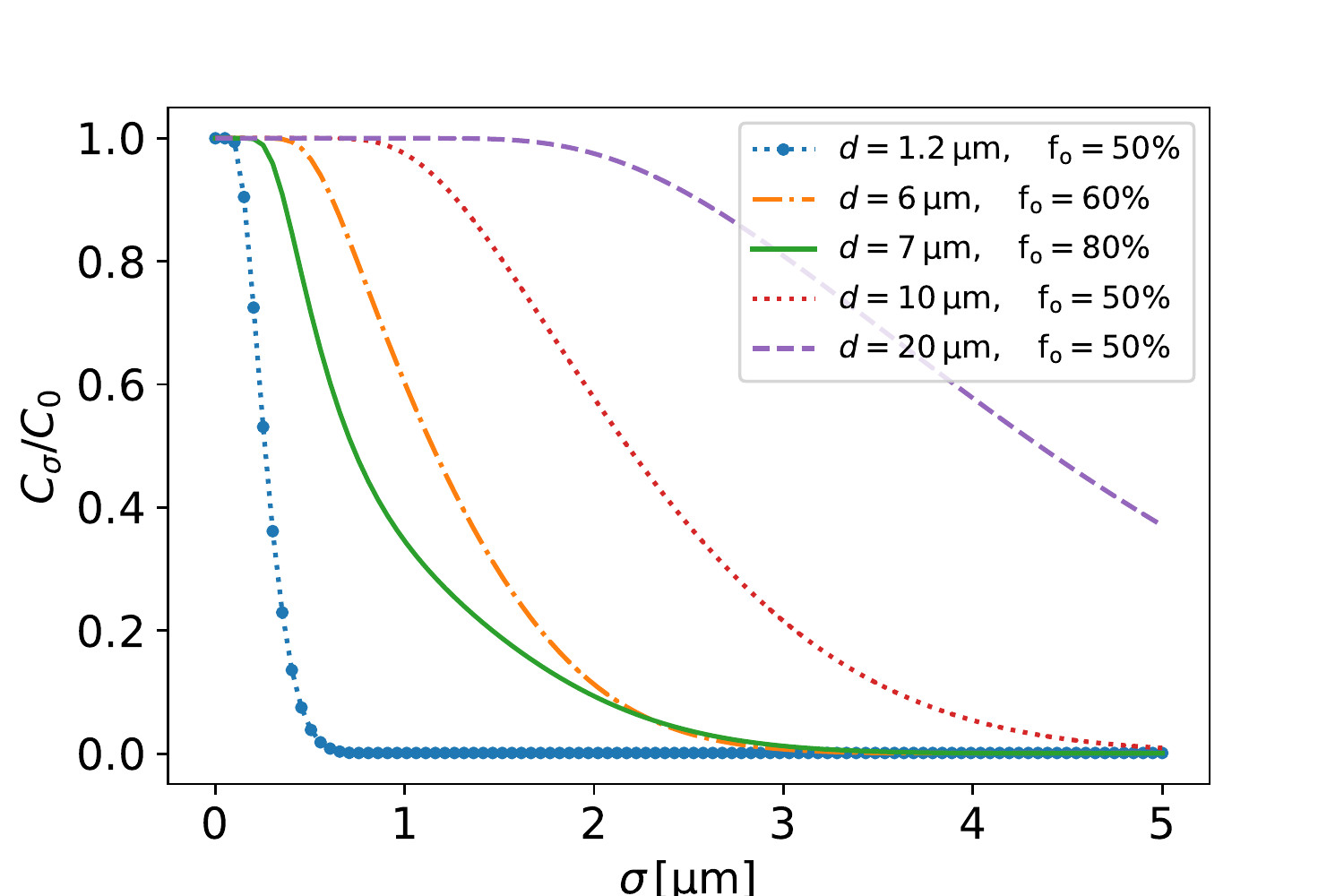}
\caption{\label{fig:psf_smearing} Plot of the ratio $C_{\sigma}/C_0$, obtained with equation \ref{eqn:smeared_function} for some representative combinations of periods and open fractions. The parameter $C_0$ is the original contrast of the periodic pattern, while $C_{\sigma}$ is the contrast after smearing with a resolution $\sigma$ (as defined in equation \ref{eqn:contrast}). The ratio is largely independent of the choice of initial contrast $C_0$.}
\end{figure}

The total effective resolution is therefore given by:
\begin{equation}
\sigma= \sqrt{\sigma_{\rm{INT}}^2 + \sigma_{\rm{MSL}}^2 + \sigma_{\rm{DIV}}^2},
\label{eqn:total_resolution}
\end{equation}
and the signal, including smearing effects, is obtained by convolving the square wave $I(t)$ with a Gaussian function. The integral $I_{\sigma} (t) = \int \exp \left[ -(t-s)^2 /(2 \sigma^2) \right] I(s) \, \rm{d}s$ is easily performed and reads:
\begin{eqnarray}
I_{\sigma}(t) = \sum_{k=-\infty} ^{\infty} B \left[ \left(1-A\right) \erf \left( \frac{d(f_o +k) - t + t_0}{\sqrt{2} \sigma} \right) + \right. \\ + \left.   A \, \erf \left( \frac{d(1 +k) - t + t_0}{\sqrt{2} \sigma} \right) - \erf \left( \frac{kd - t + t_0}{\sqrt{2} \sigma} \right) \right], \nonumber
\label{eqn:smeared_function}
\end{eqnarray} 
where the sum can be truncated to $|k| \leqslant 2$ given that $\sigma \ll d$ is expected for successful detection of the periodic signal. All overall constants are absorbed into the parameter $B$, and $t_0$ accounts for a phase shift of the periodic signal. The measured contrast $C_{\sigma}$ follows from the usual definition: 
\begin{equation}
C_{\sigma}=\frac{I_{\rm{max}} - I_{\rm{min}}}{I_{\rm{max}} + I_{\rm{min}}},
\label{eqn:contrast}
\end{equation}
where $I_{\rm{max}}$ ($I_{\rm{min}}$) is the absolute maximum (minimum) of the $I_\sigma(t)$ function.
It is worth noting that in equation \ref{eqn:smeared_function} the parameters $\sigma$ and $A$ (or $C_0$) are strongly correlated. Typically a good fit to the data is obtained with different combinations of these two parameters. Intuitively a poorer resolution can mimic the effects of a lower initial contrast and vice versa. The ambiguity is eliminated by suitable constraints on the fit parameters, consistent with a realistic estimate of $C_0$ via Monte Carlo simulation. The analytical expression \ref{eqn:smeared_function} also allows to conveniently estimate the required resolution to detect a given periodic pattern. Several examples are shown in figure \ref{fig:psf_smearing}, which suggests that a resolution significantly better than $1 \micron$ is required to detect the $6 \micron$ positron interference pattern with sizeable contrast. Nuclear emulsions are among the few detectors capable of this resolution.

\begin{figure}[htbp]
\centering 
\includegraphics[width=.49\textwidth]{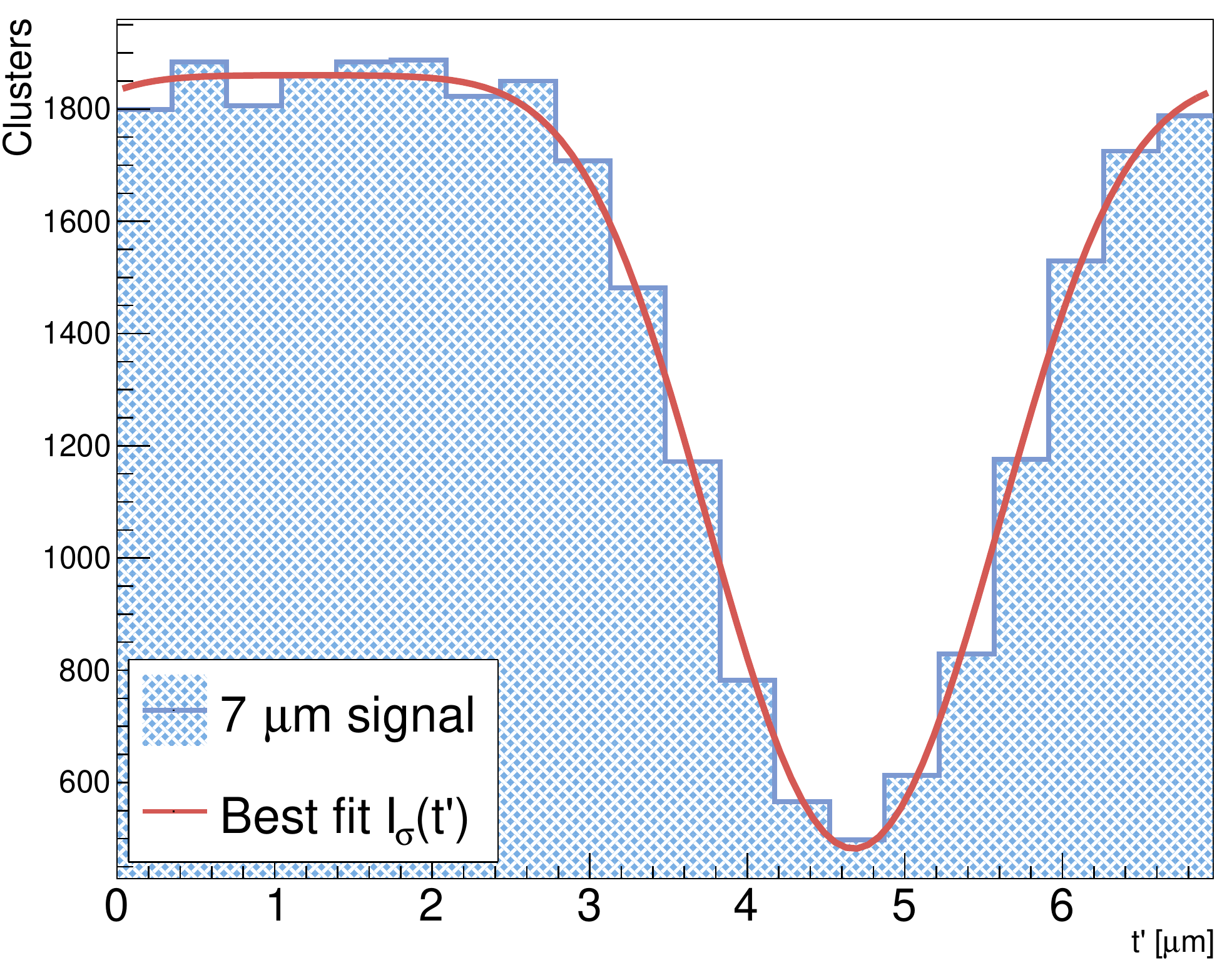}
%\qquad
\includegraphics[width=.47\textwidth]{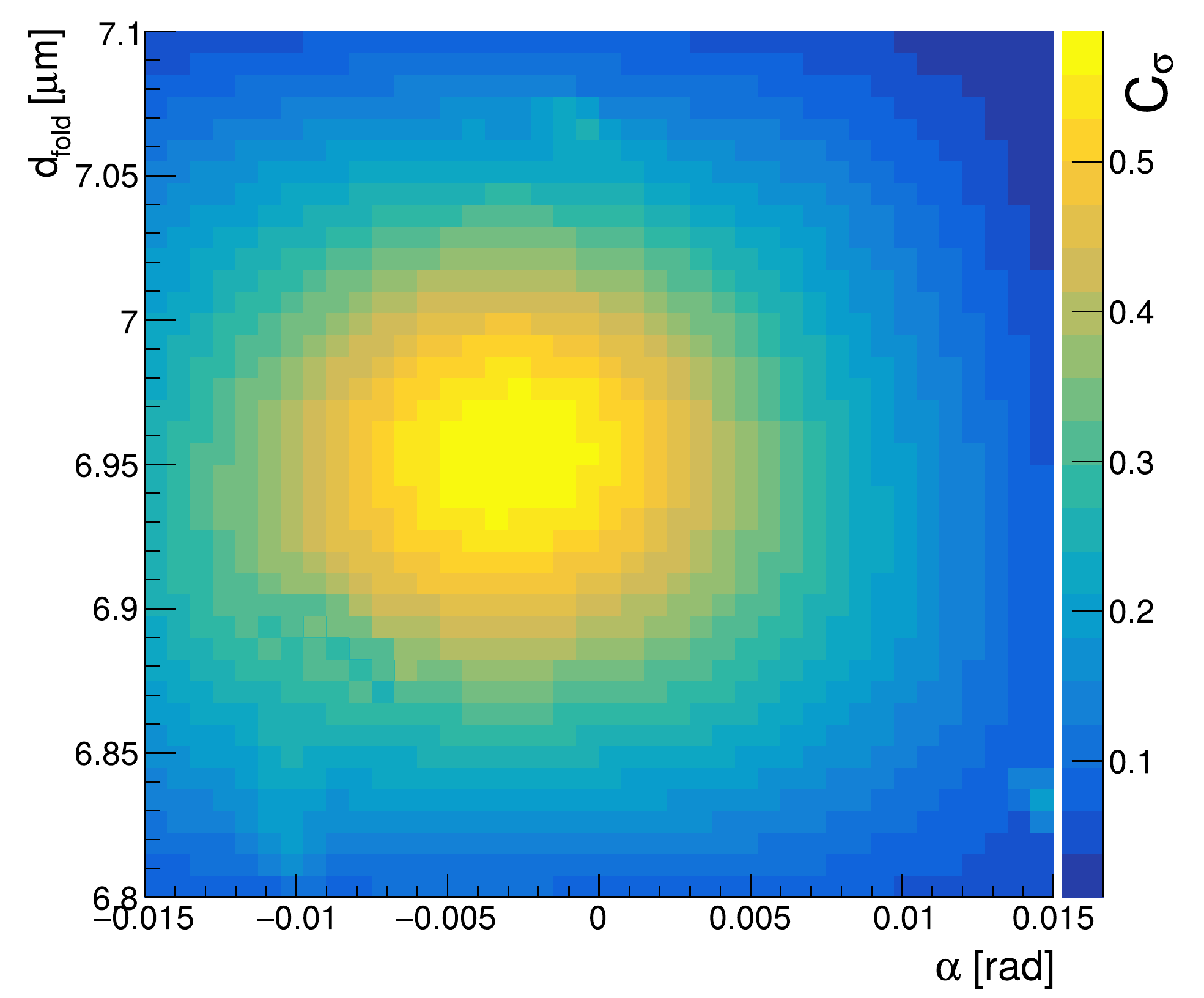}
\caption{\label{fig:blindsearch} Left: example of a fit of the distribution of the grains for the same view of Grating A exposure displayed in figure \ref{fig:analysis_fromview}. The measured contrast is approximately $50 \%$. Right: $C_{\sigma}$ as a function of $(\alpha,d_{\rm{fold}})$.}
\end{figure}

Equation \ref{eqn:smeared_function} should fit the observed intensity profile, however, to enhance the visibility of the pattern, it is convenient to \emph{fold} the data over one single period $d_{\mathrm{fold}}$, by constructing a histogram of $t' =  t \, \mathrm{mod} \, d_{\rm{fold}}$. A fit of the resulting histogram with the equation \ref{eqn:smeared_function} is then performed with the constraint $d=d_{\rm{fold}}$. An example is shown in figure \ref{fig:blindsearch}. 

The goal of the analysis is then to find the optimal combination of parameters $(  \alpha^* , d_{\rm{fold}}^*  )$ that maximize $C_{\sigma}$, therefore $d_{\rm{fold}}^*$ is the best estimate of the signal periodicity.
A two-parameter blind search is performed on a coarse-grained grid; a representative example is shown in figure \ref{fig:blindsearch}, where $C_{\sigma}$ is plotted as a function of $(\alpha,d_{\rm{fold}})$. The optimal values $(\alpha^*,d_{\rm{fold}}^*)$ are then determined by means of a standard minimization algorithm. The described analysis is performed independently on each \emph{view} of the analyzed emulsion surface.

An alternative method based on the so-called Rayleigh test \cite{mardia} was applied. This approach is effective in finding periodicities in unbinned data even with low statistics, and does not depend on a specific model for the shape of the expected signal. The test statistic, $R$, is defined as \cite{mardia} 
\begin{equation}
R(\alpha,d_{\rm{fold}}) =\left|\frac{1}{n}\sum_{j=1}^{n}  \exp{\left(i \frac{2\pi t_j(\alpha)}{d_{\rm{fold}}}\right)}\right|.
\label{eqn:rayleigh}
\end{equation}
where $t_j(\alpha)$ labels the rotated coordinate of the $j$-th cluster in a view. For large $n$, the quantity $2nR^2$ is distributed as $\chi^2$ with two degrees of freedom \cite{mardia}, which can be used to evaluate the significance of a periodic signal from the hypothesis of uniformly distributed points. Therefore, in full analogy with the previous method, a two parameter search is made to find the optimal values ($\alpha^*$,$d_{\rm{fold}}^*$) maximizing $R$.
This method yields compatible results with respect to the previous technique in finding the optimal angle and period (see discussion and examples in section \ref{sec:expresultsA}). %The confidence of the periodic pattern is obvious as the maximum value of $2nR^2=3017$ is significantly large with respect to, for example, $\chi^2=9.2$ for $p$-value=0.01.
Finally, it is worth noting that if the data from emulsion scans are used to measure length scales, such as the grating periods, the systematic error on the conversion factor between image pixels and actual position on the emulsion surface must be taken into account. This constant is estimated by moving the horizontal stage a few hundreds of microns and comparing the displacement of a pattern of clusters in microns (as it is measured by the stage encoder) and in pixels (as it is observed by the camera). We have estimated an uncertainty of approximately $0.8 \%$ on the measurement of this parameter.

\section{Results}
\label{sec:expresults}

\subsection{Grating A exposure}
\label{sec:expresultsA}
Exposure of Grating A was carried out at the positron energy $E_{A} = 12 \, \rm{keV}$. Transport through the multilayer of gold and silicon nitride (Si$_3$N$_4$) is simulated, to extract the energy distribution of the transmitted positrons, which will be degraded in energy. This distribution is then multiplied by a function $T(E)$, defined as the the probability for a positron of energy $E$ to be transmitted through the emulsion protective layer. The function $T(E)$ was known from previous measurements \cite{emulsions_old}. The integral of the resulting distribution allows one to calculate the fraction $F_C$ of positrons impinging on the active emulsion surface having traversed the closed part of the grating. The corresponding fraction for the open part of the grating is simply $T(E_{A})$. Therefore, the estimated contrast is $C_0 = \left[ T(E_{A}) - F_C  \right] / \left[ T(E_{A}) + F_C \right]$. Accounting for the errors on the protective layer thickness, a contrast range $0.75 < C_0< 0.85$ for $C_0$ was estimated. 

The technique outlined in section \ref{sec:techniques} was applied systematically to all the views in a region containing the grating membrane, to extract the parameters $(C_{\sigma},\alpha^*, d_{\rm{fold}}^*, \sigma,f_o)$, searching for the expected periodicity of $7 \micron$. The $1.2 \micron$ periodicity pattern is instead undetectable with the present experimental setup, as it will be explained.

\begin{figure}[htbp]
\centering 
\includegraphics[width=.45\textwidth]{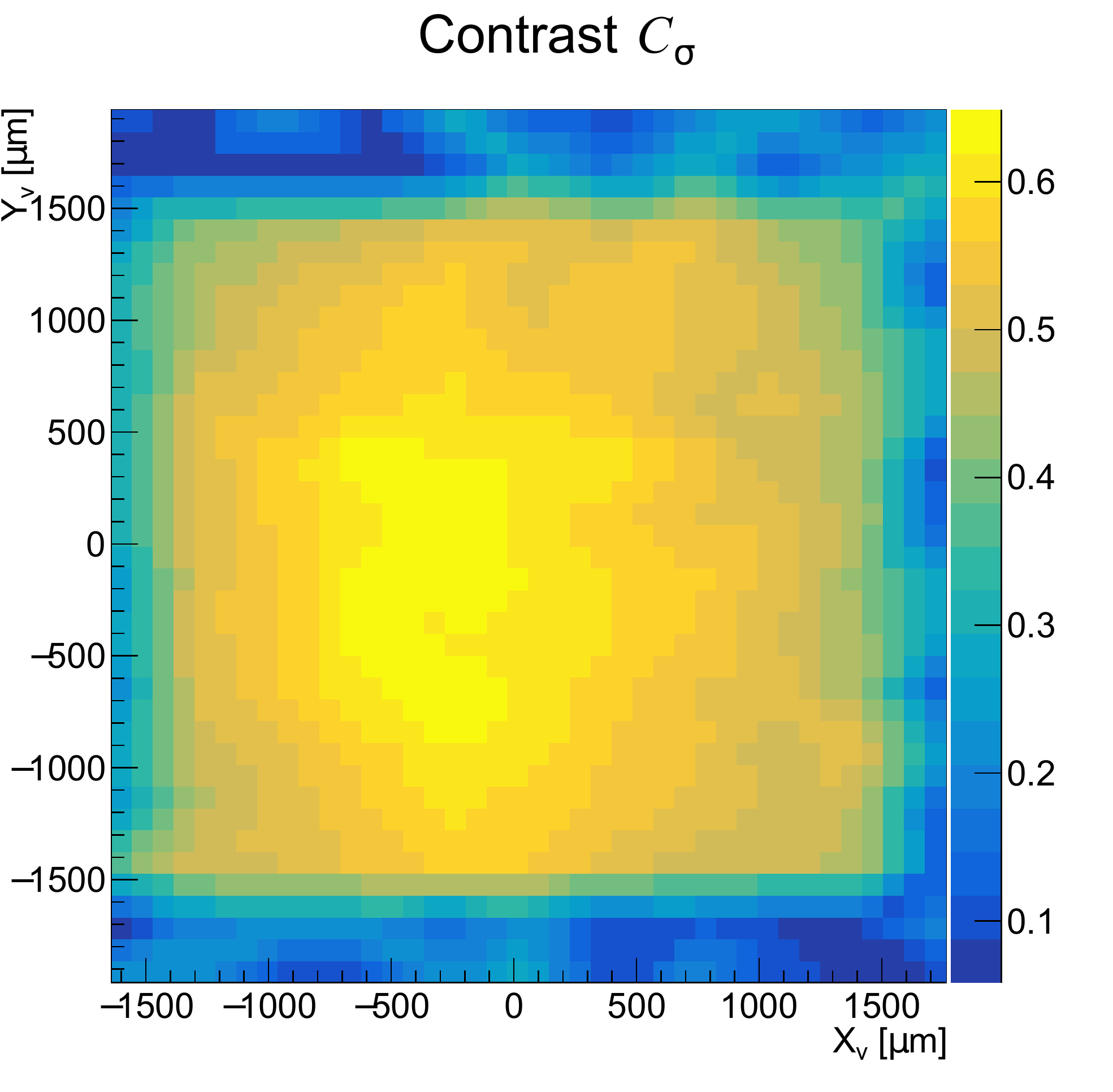}
\qquad
\includegraphics[width=.45\textwidth]{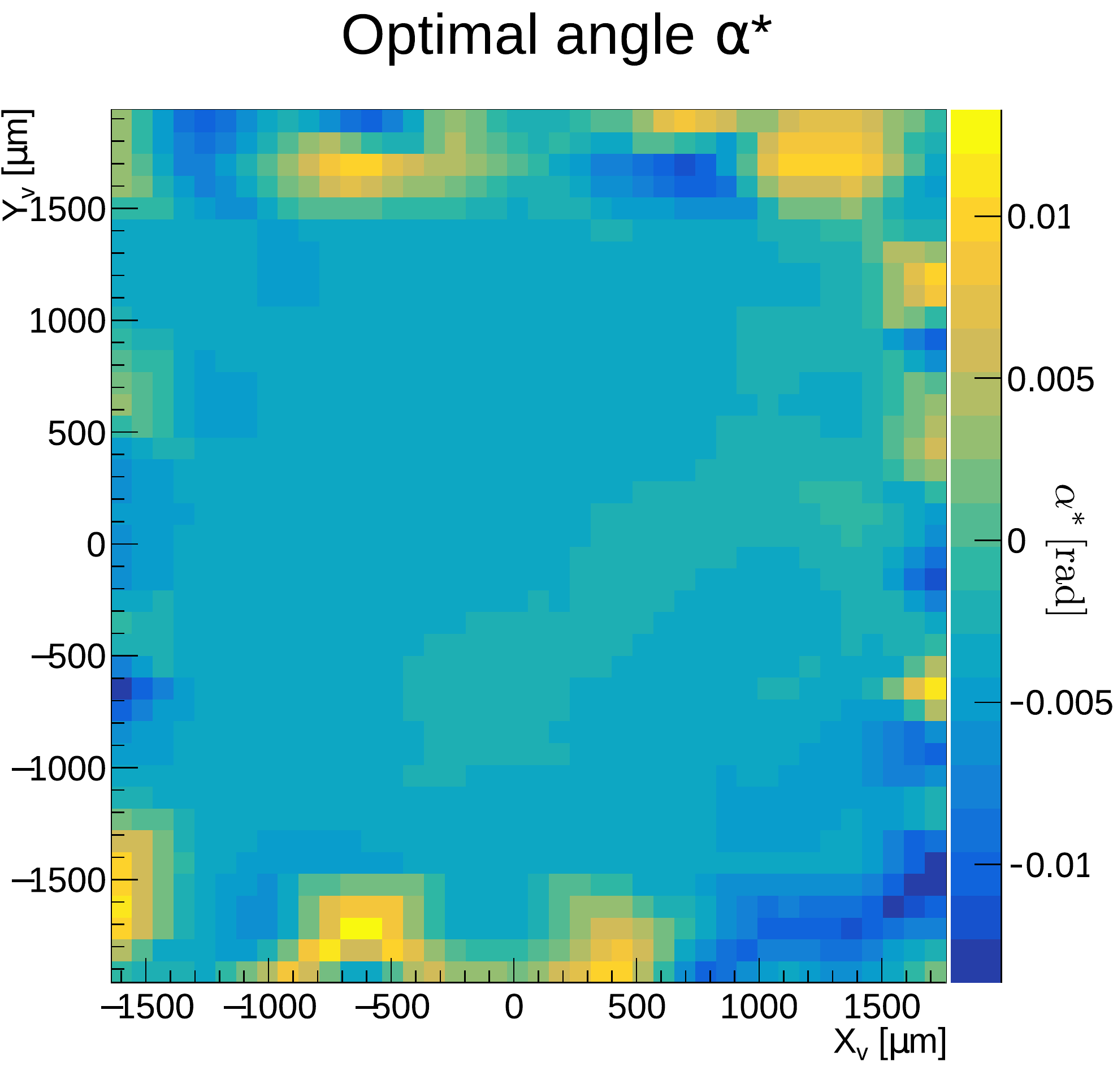}
\caption{\label{fig:expA_maps}Maps of the contrast $C_{\sigma}$ (left) and of the optimal angle $\alpha^*$ (right), as a function of the coordinates of the geometrical center of the views $(\rm{X}_v , \rm{Y}_v)$.}
\end{figure}

Figure \ref{fig:expA_maps} shows the  maps of the contrast $C_{\sigma}$ and the angle $\alpha^*$ as a function of the coordinates of the geometrical center of the view on the emulsion surface $(\rm{X}_{\rm{v}} , \rm{Y}_{\rm{v}})$.

\noindent The square shape of the membrane is clearly visible in the contrast plot (left panel of figure \ref{fig:expA_maps}). The contrast modulation within the surface is due to several factors, as for example the intensity fall-off of the beam ($\mathrm{FWHM} \approx 2.3 \, \rm{mm}$) that reduces the signal-to-noise ratio near the edge of the membrane. Misalignments between the membrane and the emulsion plane, or correlations between transverse position within the beam spot and the angular spread of incoming positrons could also be responsible for a position-dependent resolution effect. Near the boundary of the membrane region contrast does not drop abruptly as the edge of the views is not necessarily aligned with the edges of the membrane itself.

\noindent The right panel of figure \ref{fig:expA_maps} shows an equally coherent result for the angle $\alpha^*$ as  expected for a genuine periodic signal. The small spread of the angle (see table \ref{table:results_A}) is also a good indication that deformations of the emulsion after the development procedure or unwanted rotations of the microscope during its automated motions are small.

\begin{figure}[htbp]
\centering 
\includegraphics[width=.45\textwidth]{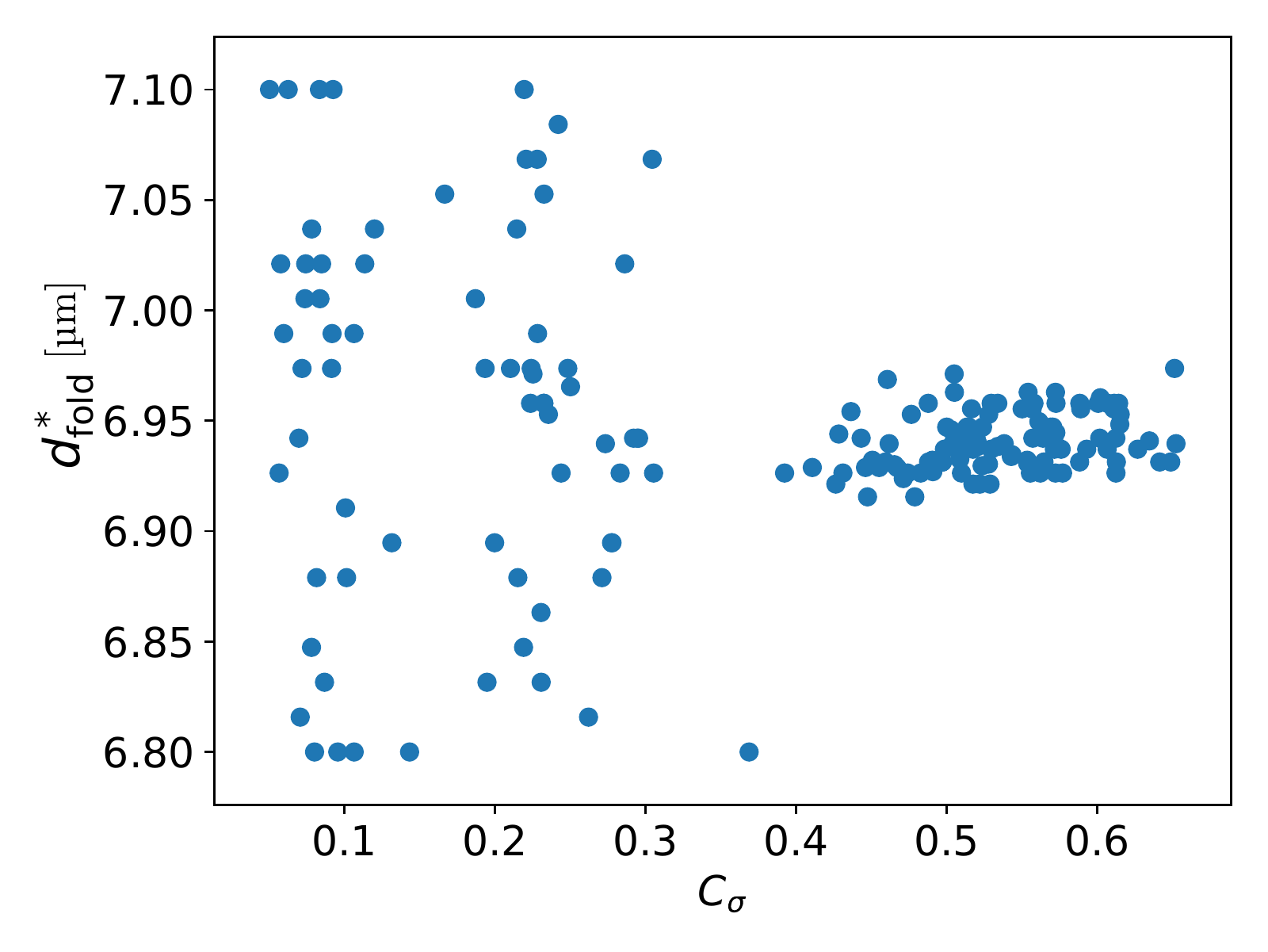}
\qquad
\includegraphics[width=.45\textwidth]{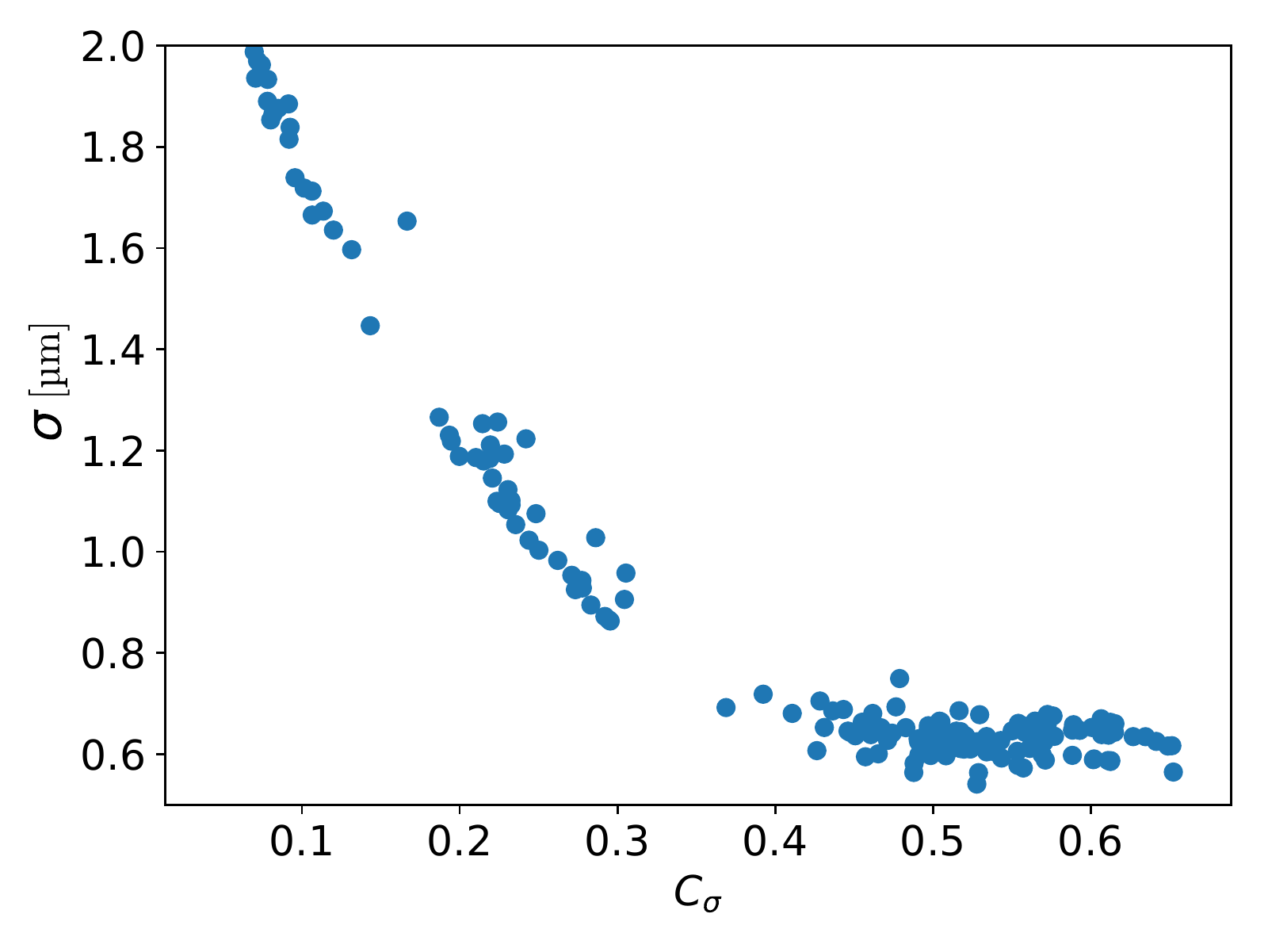}
\caption{\label{fig:expA_scatter} Scatter plots of the optimal period $d^*$ (left) and resolution $\sigma$ (right), versus the contrast $C_{\sigma}$ }
\end{figure}

In figure \ref{fig:expA_scatter} the scatter plot of the periodicity (left) and resolution (right) as a function of the contrast are shown for each view. In particular, in the left panel a coherent periodic signal is evident in a portion of the region analyzed: points with contrast $C_{\sigma} > 40 \%$ belong to views which fall entirely inside the membrane region, where fit results are physically meaningful and reliable.
An interesting trend is apparent in the right panel of figure \ref{fig:expA_scatter}, which is explained as follows: the function $I_\sigma(t)$, defined in equation \ref{eqn:smeared_function}, formally reduces to a constant (that is, a periodic signal with $C_{\sigma}=0$) either in the limit $C_0 \to 0$ or $\sigma \to \infty$. In this analysis the parameter $C_0$ is bounded to the range obtained by Monte Carlo simulation, therefore in low contrast views a good fit is obtained for larger values of $\sigma$. A correlation between low contrast and high estimated $\sigma$ therefore arises. In analogy with the previous discussion, no correlation is present for $C_{\sigma} > 40 \%$, that is on views well within the membrane region.

The best estimates of the parameters are obtained restricting the calculation to the views which fall completely inside the membrane region: $|\rm{X}_{\rm{v}}| < 1 \, \rm{mm}$ and $|\rm{Y}_{\rm{v}}| < 1 \,  \rm{mm}$. Results are summarized in table \ref{table:results_A}.

\begin{table}[htbp]
\centering
\smallskip
\begin{tabular}{|c|c|c|}
\hline
& Period Folding & Rayleigh test \\
\hline
$d_{\rm{fold}}^*$ & $(6.944  \pm 0.002_{\rm{stat}} \pm 0.050_{\rm{syst}} ) \micron$ &  $(6.942  \pm 0.002_{\rm{stat}} \pm 0.050_{\rm{syst}} ) \micron$   \\
$\alpha^*$ & $(-0.0034  \pm 0.0002  ) \, \rm{rad}$  & $(-0.0036  \pm 0.0002 ) \, \rm{rad}$ \\
$C_0$ & $ (75.5 \pm 0.2) \% $  & -\\
$C_{\sigma}$ & $(56.2  \pm 0.5 )\%$  & -\\
$f_o$  & $ (75.5 \pm 0.2) \%$ & - \\
$\sigma$ & $ (0.630 \pm 0.004_{\rm{stat}} \pm 0.005_{\rm{sist}}) \micron$ &- \\
\hline
\end{tabular}
\caption{\label{table:results_A} Best estimate of the relevant parameters for the exposure of Grating A. Results for the optimal angle and period found by the folding and Rayleigh test methods are fully compatible. Systematic errors stems from the calibration of the optical imaging system, as discussed in section \ref{sec:techniques}.}
\end{table}

\begin{figure}[htbp]
\centering 
\includegraphics[width=.99\textwidth]{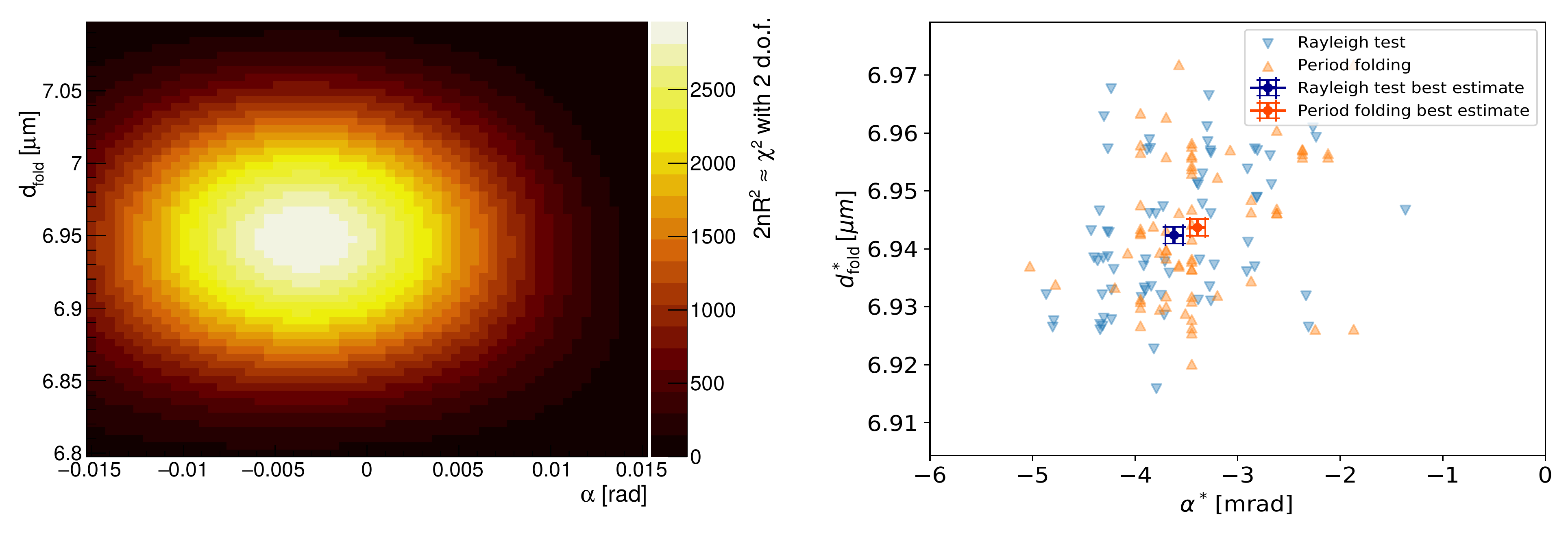}
\caption{\label{fig:expA_comparison} Left: Plot of the function $2 nR^2(\alpha, d_{\rm{fold}})$, defined from equation \ref{eqn:rayleigh}, for the same view shown in figure \ref{fig:blindsearch}. Right: scatter plot of the optimal values $(\alpha^* ,d_{\rm{fold}}^* )$ for the views in the membrane region for both the period folding and the Rayleigh test methods.}
\end{figure}

We have also applied to the same set of data the Rayleigh test method described in section \ref{sec:matmethods} as a consistency check. The two parameter search in the $(\alpha,d_{\rm{fold}})$ plane has been found to yield compatible results on single views (compare the left panel of figure \ref{fig:expA_comparison} with the right panel of figure \ref{fig:blindsearch}). A scatter plot of the optimal values $(\alpha^* ,d_{\rm{fold}}^* )$ found on the set of views within the membrane with both methods is shown in the right panel of figure \ref{fig:expA_comparison}. The best estimates of the two parameters produced by the methods are compatible within their errors.

A periodic pattern with an average contrast of approximately $56 \%$ was detected, with features consistent with the theoretical model of equation \ref{eqn:smeared_function}. The measured period and open fraction are compatible with the nominal specifications provided by the manufacturer. The period in particular was measured with two different methods, which yield fully compatible results.

It is worth noting that the estimated overall resolution of $0.63 \micron$ reported in table \ref{table:results_A} is fully compatible with expectations, as described in section \ref{sec:techniques}, and such a value prevents the detectability of the $1.2 \micron$ pattern. This also means that with a more optimized setup, with smaller membrane-to-emulsion distance and possibly a thinner protective layer, the detection of periodic patterns down to the $\approx 1 \micron$ length scale in a genuine interferometer is likely possible and will be investigated in the future. 

\subsection{Grating B exposure}
\label{sec:expresultsB}

\begin{figure}[htbp]
\centering 
\includegraphics[width=.45\textwidth,trim={0 -1.5cm 0 1cm},clip]{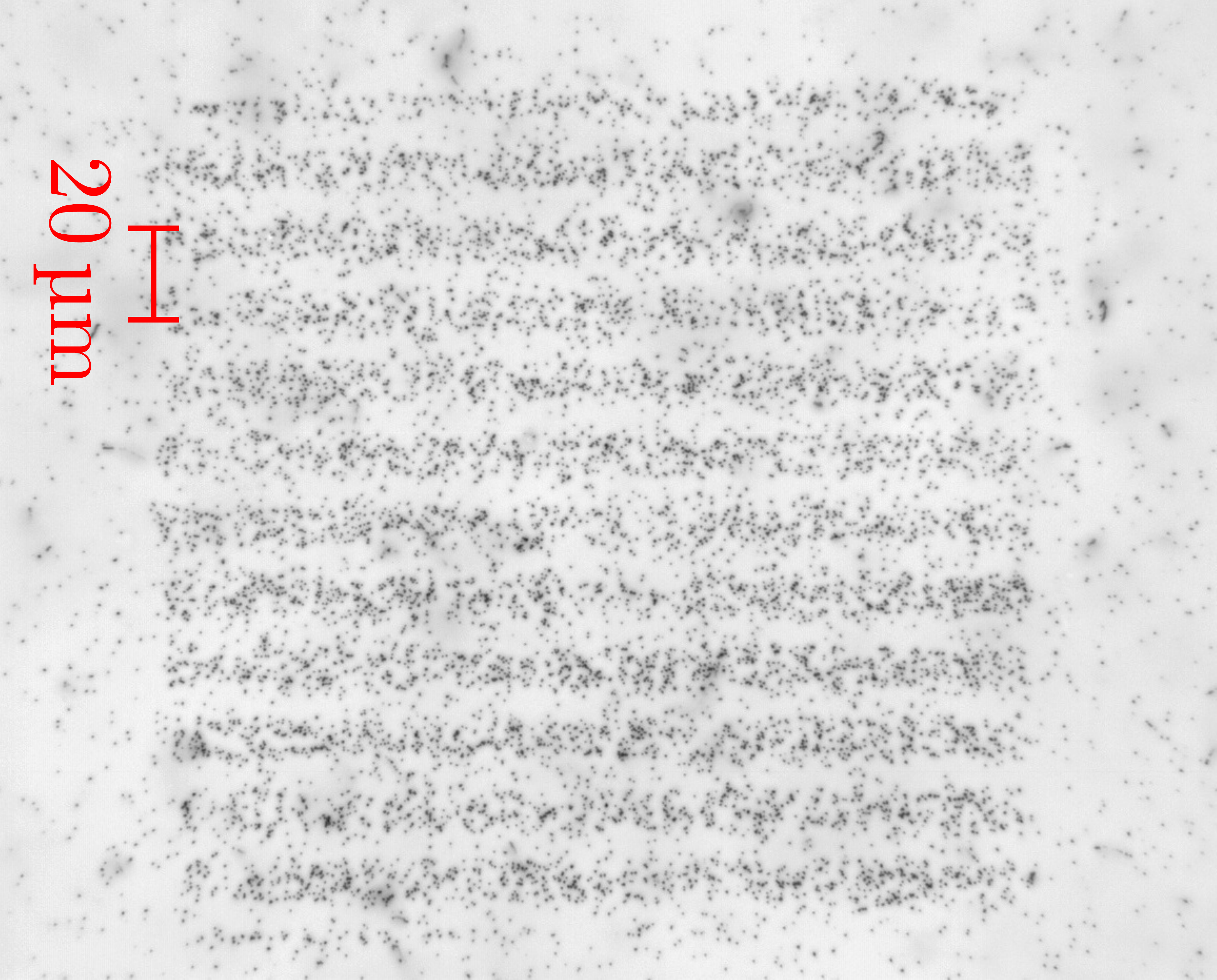}
\qquad
\includegraphics[width=.45\textwidth,trim={0 0 0 0.0},clip]{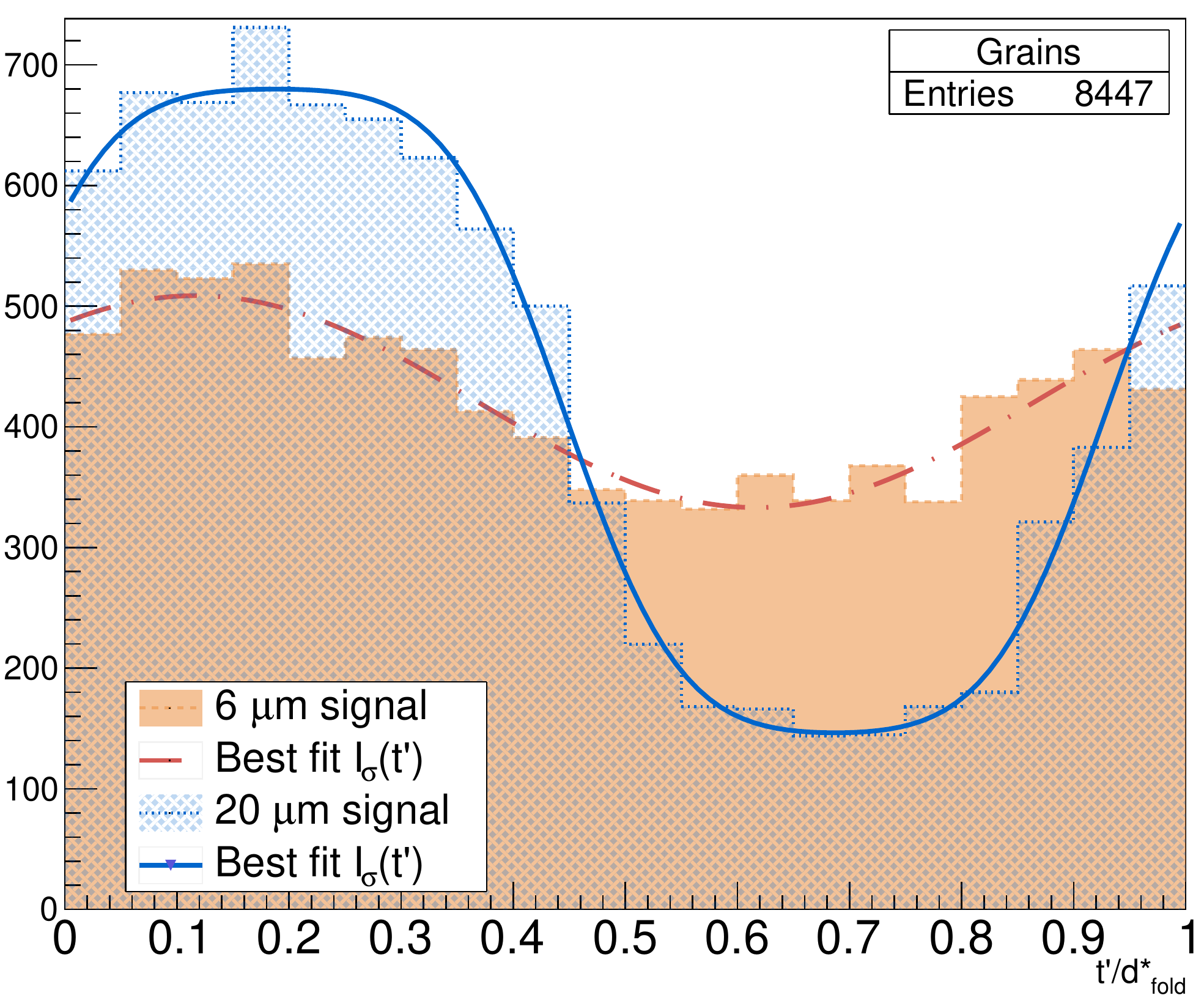}
\caption{\label{fig:expB_images} Left: raw microscope image from a single view containing one of the open regions of the diffraction grating (see figure \ref{fig:gratings}). The $20 \micron$ support structure is clearly visible, whereas the $6 \micron$ periodicity in the orthogonal direction cannot be distinguished. Right: folded signal of a single view for both periodicities at their optimal values of period and angle, compared with the best fit with equation \ref{eqn:smeared_function}.}
\end{figure}

For the exposure of Grating B a positron energy of $E_B = 10 \, \rm{keV}$ was selected, to minimize the amount of positrons transmitted by the closed parts of the membrane, which is significantly thinner than that of Grating A. As stated in section \ref{sec:matmethods}, Grating B is composed of several disconnected regions, therefore an extensive analysis of the whole surface was not meaningful. A total of 6 views displaying a clear periodic signal with the $20 \micron$ periodicity were considered; a typical raw microscope frame from such regions is shown in figure \ref{fig:expB_images} (left panel).

Each contains two detectable periodic patterns, a feature which can be exploited to improve the results of the analysis. Figure \ref{fig:psf_smearing} suggests that the contrast of the $20 \micron$ pattern is essentially independent of the smearing resolution $\sigma$ up to $\sigma \approx 2 \micron$, therefore $C_{\sigma}(20 \micron) \approx C_0(20 \micron)$. On the contrary, the $6 \micron$ signal is highly sensitive to this parameter. Furthermore, the parameters $C_0$ and $\sigma$ are independent of the periodicities of the patterns and in ideal conditions should have the same value for both periodic signals. First the Rayleigh test method was applied to find the optimal values $(\alpha^* ,d_{\rm{fold}}^* )$, then a fit with equation \ref{eqn:smeared_function} was performed on the $20 \micron$ signal to extract a reliable estimate of $C_0 = \overline{C}_0 \pm \sigma_{\overline{C}_0}$, where the uncertainly $\sigma_{\overline{C}_0}$ is obtained from the fit. Subsequently, equation \ref{eqn:smeared_function} is fit to the $6 \micron$ pattern with the constraint $|C_0 -\overline{C}_0 |<\sigma_{\overline{C}_0}$ to obtain an estimate of all the parameters, including $\sigma$. Finally, a fit of the $20 \micron$ pattern is performed again with suitable constraints on both $C_0$ and $\sigma$. The final results after this procedure averaged over the views are summarized in table \ref{table:results_B}. Folded signals at the optimal angles and period are shown for a representative view in the right panel of figure \ref{fig:expB_images}, and compared to the best fit with equation \ref{eqn:smeared_function}.

\begin{table}[htbp]
\centering
\smallskip
\begin{tabular}{|c|c|c|}
\hline
& $20 \micron$ pattern & $6 \micron $ pattern \\
\hline
$d_{\rm{fold}}^*$ & $(19.2  \pm 0.1_{\rm{stat}} \pm 0.2_{\rm{syst}}  ) \micron$ &  $(5.93  \pm 0.02_{\rm{stat}} \pm 0.05_{\rm{syst}} ) \micron$   \\
$C_{\sigma}$ & $(60  \pm 2 )\%$  & $(17  \pm 1 )\%$ \\
$C_0$ & $(60  \pm 2 )\%$  & $(60  \pm 2 )\%$ \\
$f_o$  & $ (53 \pm 1) \%$ & $(57 \pm 1) \%$ \\
$\sigma$ & $(1.58 \pm 0.05) \micron $& $(1.52 \pm 0.05) \micron$ \\
\hline
\end{tabular}
\caption{\label{table:results_B} Best estimate of the relevant parameters for the exposure of Grating B, for the two periodic patterns with nominal periodicities of $20 \micron$ and $6 \micron$.}
\end{table}

The angle $\alpha^*$ is not reported, since for this grating the disconnected regions do not have the same orientation. The procedure yields self-consistent results between the two patterns as the values of $C_0$ and $\sigma$ are compatible. The poorer effective resolution obtained in this case is due to the increased membrane-to-emulsion distance in comparison to Grating A. The distance $\Delta X_B \approx 0.3 \, \rm{mm}$, introduced in equation \ref{eqn:total_resolution} with the estimated values of $\sigma_\theta$ and $\sigma_{\rm{MSL}}$ reported in section \ref{sec:techniques}, yields a predicted total resolution $\sigma_B \approx 1.5 \micron$, in good qualitative agreement with the measured value.  

\subsection{Long exposure in High Vacuum}
\label{sec:expresultsvac}

As anticipated, matter-wave interferometry with low-intensity antimatter beams can require very long exposure times. It was reported in the past that emulsion films undergo an increase of background  when kept in High Vacuum (HV) conditions (namely pressures lower than $10^{-6} \, \rm{mbar}$) for several days \cite{emulsions_in_vacuum}. 
However, our detector features a new gel composition, therefore we investigated the possible background increase repeating the exposure of Grating A at the same positron energy, after the emulsion was kept in the vacuum chamber at a pressure of the order of $10^{-6}-10^{-7} \, \rm{mbar}$ for $1$ week.

\begin{figure}[htbp]
\centering 
\includegraphics[width=.445\textwidth]{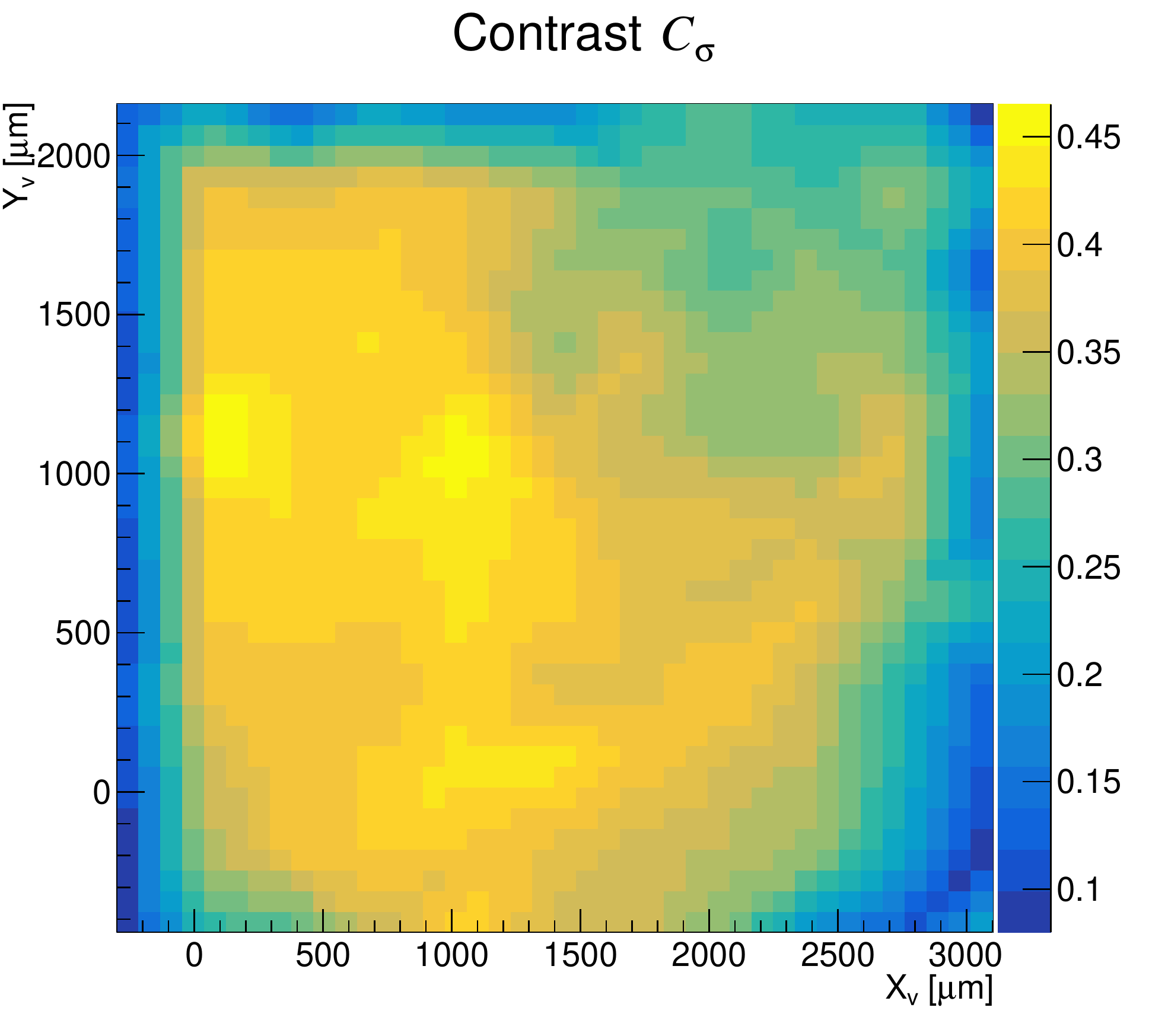}
\qquad
\includegraphics[width=.45\textwidth,trim={-0.8cm 0 0 0.0},clip]{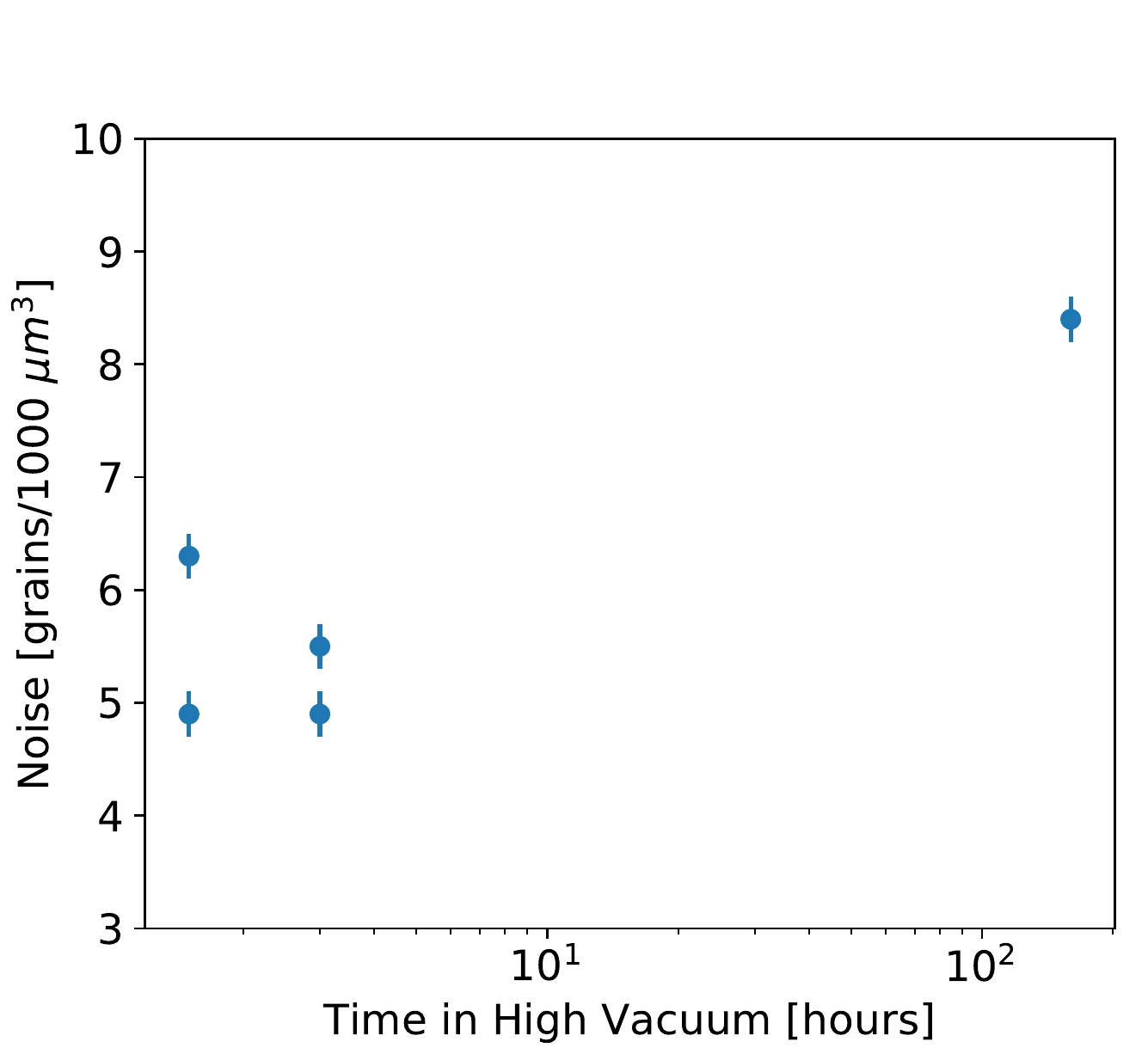}
\caption{\label{fig:long_exposure} Left: Contrast map covering the membrane surface, for an exposure performed after $1$ week in High Vacuum (HV) conditions, at a pressure $\gtrsim 10^{-7} \, \rm{mbar}$ .  Right: Density of grains measured at a distance of about $4 \, \rm{mm}$ from the center of the beam spot, for different emulsion samples and positron exposure times as a function of the total time spent in HV.}
\end{figure}

The contrast map in the left panel of figure \ref{fig:long_exposure} confirms that the periodic structure is still well resolved. A direct comparison with the previous results would be misleading since the membrane/emulsion positioning was altered. The background measurements are displayed in the right panel of figure \ref{fig:long_exposure}. Increasing the exposure time in high vacuum by two orders of magnitude induced an increase in background of a factor around $1.5$ with respect to the average of the other four observations. This indicates that the background level is not directly proportional to the time spent in vacuum, therefore long exposure times can effectively increase the signal-to-noise ratio.

\section{Conclusions}
\label{sec:conclusions}
 
The purpose of this study was to demonstrate the detectability of micrometric-scale periodic patterns with emulsion detectors, in view of future applications to positron interferometry. 
We exposed two different diffraction gratings featuring nominal periodicities of $6,7$ and $20 \micron$ to a positron beam with $10$ or $12 \, \rm{keV}$ energy. 
For the employed setup, a study based on a realistic physical model suggested that in the context of interferometry resolutions of the order of $\sigma \lesssim 0.4 \micron$ can be achieved for positrons in this energy range, if no spurious smearing effects are present. 
Data were analysed using two different methods, which led to fully compatible results. We successfully detected the $6,7$ and $20 \micron$ patterns with a contrast of $(17 \pm 1) \%$, $(56.2 \pm 0.5 ) \%$ and $(60 \pm 2) \%$ respectively.

The possibility to perform long-time exposures was also investigated. For interferometry experiments with antimatter beams where typical intensities are orders of magnitudes lower than standard atomic or electron beams, several days of continuous data taking could be required to collect the necessary statistics. An exposure performed after keeping the emulsion detector in high vacuum conditions for one week resulted in a background increase of a factor 1.5 with respect to typical hour-long exposures and did not prevent the detectability of the periodic pattern.

We can conclude that emulsion detectors are a promising option to carry out matter-wave interferometry with positrons.

\section{Acknowledgements}
\label{sec:ackn}

The authors would like to thank Marco Leone, Roger Haenni, Giancarlo Maero, Massimiliano Rom\`{e}, Simone Cialdi, Fabrizio Castelli and Marco Potenza for their valuable contribution.

\vspace{1cm}

%\bibliographystyle{iopart-num}
%\bibliography{bibliography}

\begin{thebibliography}{10}
\expandafter\ifx\csname url\endcsname\relax
  \def\url#1{{\tt #1}}\fi
\expandafter\ifx\csname urlprefix\endcsname\relax\def\urlprefix{URL }\fi
\providecommand{\eprint}[2][]{\url{#2}}





\bibitem{amattint1}
Sala S, Castelli F, Giammarchi M, Siccardi S and Olivares S 2015 {\em J. Phys.
  B:At. Mol. Opt. Phys.\/} {\bf 48} 195002

\bibitem{amattint2}
Sala S, Giammarchi M and Olivares S 2016 {\em Phys. Rev. A\/} {\bf 94}(3)
  033625

\bibitem{emulsions_old}
Aghion S, Ariga A, Ariga T, Bollani M, {Dei Cas} E, Ereditato A, Evans C,
  Ferragut R, Giammarchi M, Pistillo C, Rom\`{e} M, Sala S and Scampoli P 2016
  {\em J. Instrum.\/} {\bf 11} P06017

\bibitem{lau}
Lau E 1948 {\em Ann. Phys. (Berl.)\/} {\bf 437} 417--423 ISSN 1521-3889

\bibitem{talbotlau}
Clauser J~F and Li S 1994 {\em Phys. Rev. A\/} {\bf 49}(4) R2213--R2216

\bibitem{debroglie}
de Broglie L, 1925 {\em Ann. Phys. (Paris) }{\bf 3} 22 

\bibitem{aegis2}
Amsler C, Ariga A, Ariga T, Braccini S, Canali C, Ereditato A, Kawada J, Kimura
  M, Kreslo I, Pistillo C, Scampoli P and Storey J~W 2013 {\em J. Instrum.\/}
  {\bf 8} P02015

\bibitem{nature}
Aghion S {\em et~al.\/} (AEgIS) 2014 {\em Nature Commun.\/} {\bf 5} 4538

\bibitem{croninmodel}
McMorran B and Cronin A~D 2008 {\em Phys. Rev. A\/} {\bf 78}(1) 013601

\bibitem{three-grating}
Gronniger G, Barwick B and Batelaan H 2006 {\em New J. Phys.\/} {\bf 8} 224

\bibitem{emulsion_technology}
Ereditato A 2013 {\em Adv. High Energy Phys.\/} {\bf 2013} 1

\bibitem{penelope}
Bar{\'o} J, Sempau J, Fern{\'a}ndez-Varea J and Salvat F 1995 {\em Nucl. Instr.
  Meth. Phys. Res. Sec. B: Beam Interactions with Materials and Atoms\/} {\bf
  100} 31 -- 46 ISSN 0168-583X

\bibitem{pypen}
\url{http://pypenelope.sourceforge.net/} [Online; accessed 16-January-2018]

\bibitem{mardia}
Mardia M~K 1972 {\em Statistics of directional data\/} (Academic Press, New
  York)

\bibitem{emulsions_in_vacuum}
Kimura M {\em et~al.\/} 2013 {\em Nucl. Instr. Meth. Phys. Res. Sec. A:
  Accelerators, Spectrometers, Detectors and Associated Equipment\/} {\bf 732}
  325 -- 329 ISSN 0168-9002 Vienna Conference on Instrumentation 2013

\end{thebibliography}

\providecommand{\newblock}{}

\end{document}